\documentclass[pdflatex, sn-mathphys-ay]{sn-jnl}

\usepackage{graphicx}%
\usepackage{multirow}%
\usepackage{amsmath,amssymb,amsfonts}%
\usepackage{amsthm}%
\usepackage{mathrsfs}%
\usepackage[title]{appendix}%
\usepackage{xcolor}%
\usepackage{textcomp}%
\usepackage{manyfoot}%
\usepackage{booktabs}%
\usepackage{algorithm}%
\usepackage{algorithmicx}%
\usepackage{algpseudocode}%
\usepackage{listings}%
\usepackage{caption}

\usepackage{epstopdf}
\DeclareGraphicsExtensions{.pdf,.eps,.png,.jpg}

\theoremstyle{thmstyleone}%
\theoremstyle{thmstyletwo}%

\theoremstyle{thmstylethree}%

\begin{document}

\title[Article Title]{BugRepro: Enhancing Android Bug Reproduction with
Domain-Specific Knowledge Integration}


\author[1]{\fnm{Hongrong} \sur{Yin}}\email{hongrongyin@proton.me}
\author[1]{\fnm{Jinhong} \sur{Huang}}\email{djywithhjh@gmail.com}
\author[1]{\fnm{Yao} \sur{Li}}\email{liyao@must.edu.mo}
\author[2]{\fnm{Yunwei} \sur{Dong}}\email{yunweidong@nwpu.edu.cn}
\author*[1]{\fnm{Tao} \sur{Zhang}}\email{tazhang@must.edu.mo}

\affil[1]{\orgdiv{Macau University of Science and Technology}, \orgaddress{\state{MACAU}, \country{China}}}
\affil[2]{\orgdiv{Northwestern Polytechnical University}, \orgaddress{\state{Xi'an}, \country{China}}}




\abstract{Mobile application development is a fast-paced process where maintaining high-quality user experiences is crucial. Bug reproduction, a key aspect of maintaining app quality, often faces significant challenges. Specifically, when descriptions in bug reports are ambiguous or difficult to comprehend, current approaches fail to extract accurate information. Moreover, modern applications exhibit inherent complexity with multiple pages and diverse functionalities, making it challenging for existing methods to map the relevant information in bug reports to the corresponding UI elements that need to be manipulated. To address these challenges, we propose BugRepro, a novel technique that integrates domain-specific knowledge to enhance the accuracy and efficiency of bug reproduction. BugRepro adopts a Retrieval-Augmented Generation (RAG) approach. It retrieves similar bug reports along with their corresponding steps to reproduce (S2R) entities from an example-rich RAG document. In addition, BugRepro explores the graphical user interface (GUI) of the app and extracts transition graphs from the user interface to incorporate app-specific knowledge to guide large language models (LLMs) in their exploration process. Our experiments demonstrate that BugRepro significantly outperforms two state-of-the-art methods (ReCDroid and AdbGPT). For S2R entity extraction accuracy, it achieves a 7.57 to 28.89 percentage point increase over prior methods. For the bug reproduction success rate, the improvement reaches 74.55\% and 152.63\%. In reproduction efficiency, the gains are 0.72\% and 76.68\%.}




\maketitle

\section{Introduction}\label{sec1}

Mobile applications have become an essential part of daily life, with over 4 million apps available on Google Play and the Apple App Store as of 14 December 2024~\citep{42matters_stats}.
As the market for mobile apps continues to grow, ensuring a seamless user experience has become a critical priority for developers striving to remain competitive. 
A key aspect of maintaining app quality is addressing the issues reported by users, typically through bug reports. These reports often include S2R entities, enabling developers to replicate the reported problems, identify their root causes, and implement fixes. 

However, these reproduction steps are frequently unclear, incomplete, or inaccurate. As a result, manually reproducing bugs can be time consuming and resource intensive, further delaying the process of bug resolution~\citep{rahman2022works,aranda2009secret,just2018comparing}. To address this challenge, many automated solutions have been proposed to assist in bug reproduction. These methods typically leverage Natural Language Processing (NLP) techniques to extract structured S2R entities from bug reports using predefined rules or learned models~\citep{fazzini2018automatically,zhao2019recdroid,zhao2022recdroid+,huang2023context}. Then they leverage heuristic search algorithms to explore the app's graphical user interface (GUI) and replay the extracted S2R entities. Despite these advancements, current automated methods still face significant difficulties due to the ambiguity and vagueness of S2Rs, which poses a significant challenge to state-of-the-art NLP techniques~\citep{word2vec_2018,nogueira2019passage}. Moreover, the complexity of modern mobile apps further complicates the task of accurately replaying the extracted S2R entities~\citep{huang2023context,feng2024prompting}.
Recently, large language models (LLMs) have developed rapidly and demonstrated remarkable capabilities in understanding diverse linguistic expressions~\citep{jansen2023employing}. Building on this, Sidong Feng et al. propose AdbGPT~\citep{feng2024prompting} , which harnesses the power of LLMs to improve bug reproduction. Specifically, AdbGPT utilizes the in-context learning abilities of LLMs to extract S2R entities by learning from developer-provided few-shot examples. Then, AdbGPT iteratively asks the LLM to match S2R entities with GUI events to reproduce the bugs. 

Although prior approaches have made significant contributions to improving bug reproduction efficiency, critical limitations remain unresolved. We crystallize these challenges as follows:

\textbf{Challenge 1: Lack of Domain Understanding in Bug Report Interpretation.} Since S2Rs provided by non-expert users frequently exhibit a lack of precision and often omit essential procedural steps, thereby compromising their completeness and clarity. Prior work demonstrates LLMs' limited comprehension of Android bug reports, yielding hallucinated outputs in S2R extractions~\citep{lewis2020retrieval,wu2024faithful,huang2025survey}, making them ineffective at extracting structured S2R entities from unstructured bug reports.
While AdbGPT~\citep{feng2024prompting} attempts to address this issue by leveraging few-shot examples; however, its performance falls short due to the small number of examples (only 15) and the high cost of manually constructing them. 

\textbf{Challenge 2: Lack of Dynamic Behavior Modeling in App Reproduction.} Modern apps often involve multi-page flows and complex functionalities, making it difficult to replay S2Rs without app-specific knowledge of screen transitions and dynamic GUI states. For instance, reproducing a bug in a shopping app’s checkout process requires navigating several pages, each dependent on user actions and dynamic elements. Existing methods, including LLM-based ones, lack this contextual understanding, often resulting in failed reproductions.

This motivates our core research question: \textbf{How can we effectively leverage LLMs to serve as the ``brain'' of developers and better facilitate the reproduction of Android crashes?}

To overcome these constraints, we propose BugRepro, a novel technique that integrates bug-report-specific knowledge and app-specific knowledge to enhance the bug reproduction process. \textbf{Specifically, to overcome the first challenge,} BugRepro first employs a Retrieval-Augmented Generation (RAG)~\citep{lewis2020retrieval} approach to retrieve similar bug reports and their corresponding S2R entities from an example-rich RAG document. The retrieved knowledge serves as a reference, addressing the limitation of existing LLMs---their lack of domain understanding in bug report interpretation. This enables LLMs to extract S2R entities more accurately for the target bug report. \textbf{To overcome the second challenge,} BugRepro incorporates app-specific knowledge by exploring the app's GUI to improve the LLM-guided bug reproduction process further. When LLMs encounter challenges, such as difficulty in determining which UI component to interact with in the reproduction process, BugRepro extracts UI transition graphs(UTG) that model the changes and interactions of the app state on different screens. These graphs are then fed into the LLM to guide its decision-making, addressing the limitation of existing LLMs---their lack of dynamic behavior modeling in app reproduction. This ensures that the reproduction follows the correct sequence of actions and app state transitions.

We conduct extensive experiments on 151 real-world bug reports to evaluate the performance of BugRepro. First, we evaluate the accuracy of S2R extraction and compare BugRepro with two state-of-the-art techniques (i.e., RecDroid and AdbGPT). Experimental results show that BugRepro's extraction accuracy is 7.57 to 28.89 percentage points higher than prior methods. Then, we evaluate the effectiveness of bug replay. The experimental results show that RecDroid and AdbGPT successfully reproduced 38 and 55 bugs, while BugRepro can reproduce 96 bugs, achieving an improvement of 152.63\% and 74.55\%, respectively. Furthermore, BugRepro reduces the bug reproduction time to 124.7 seconds on average, compared to 534.9 seconds for RecDroid and 125.6 seconds for AdbGPT. Finally, ablation studies confirm that each component of BugRepro contributes to the overall performance.

The contributions of this paper are as follows:
\begin{itemize}
    \item We propose BugRepro, a novel technique that integrates LLMs with domain-specific knowledge to enhance Android bug reproduction.
    \item BugRepro effectively addresses two key challenges in Android bug reproduction: the lack of domain-specific knowledge, the complexity of modern applications with multi-page structures and diverse functionalities.
    \item We evaluate the effectiveness of BugRepro through extensive experiments, demonstrating significant improvements over two state-of-the-art methods in terms of S2R entity extraction accuracy, bug reproduction success rate, and reproduction efficiency.
\end{itemize}

The rest of the paper is organized as follows: Section~\ref{sec2}
provides an introduction to the background and motivation for this work. Section~\ref{sec3} details the analysis of our method. Section~\ref{sec4} formulates four key research questions, characterizes the experimental data and configuration, reports empirical findings with the corresponding evaluations. Section~\ref{sec5} presents the results corresponding to the research questions and analyzes them in detail. Section~\ref{sec6} discusses the limitation of our method. Section~\ref{sec7} reviews existing research in related fields. Finally, Section~\ref{sec8} concludes the paper. 

\section{Background and Motivation}\label{sec2}

\subsection{Large Language Models}\label{subsec2.1}
LLMs represent a significant advancement in natural language understanding and generation, leveraging the Transformer architecture to achieve remarkable performance across a broad spectrum of tasks. These models, including examples such as Claude3.5~\citep{anthropic2024claude35sonnet}, GPT-4~\citep{openai2023gpt4}, LLaMA~\citep{llama2024}, and DeepSeek~\citep{deepseek2024} are characterized by their billions of parameters and training on massive text corpora. Their scale and design enable capabilities like mathematical reasoning, program synthesis, and multi-step reasoning that outperform traditional benchmark models tailored to specific tasks.
The core functionality of LLMs revolves around processing prompts—task-specific instructions in natural language~\citep{liu2023pre,brown2020language}. Prompts are tokenized into words or subwords and passed through layers of the Transformer model, which employs multi-head self-attention, normalization, and feed-forward networks to comprehend input and generate contextually appropriate responses. Through this mechanism, LLMs can perform tasks without requiring task-specific fine-tuning, relying instead on prompt engineering to elicit desired behaviors.
In this work, we leverage LLMs to tackle the complex task of bug reproduction. By carefully designing prompts, we instruct the LLMs to extract S2R entities, generate actionable operations, and decode their responses to replicate bugs effectively.

However, applying LLMs to bug reproduction requires precise mapping from natural language to UI actions—a task where hallucinated outputs may occur.

\subsection{Retrieval-Augmented Generation}\label{subsec2.2}
RAG~\citep{lewis2020retrieval} is a hybrid approach that enhances the capabilities of LLMs by integrating retrieval mechanisms. Unlike standalone LLMs, which rely solely on their pre-trained knowledge, RAG retrieves relevant information from external knowledge sources, such as databases or document repositories, to improve response accuracy and relevance. This process involves two key components:

\begin{itemize}
\item Retrieval phase: A retrieval model, often based on dense vector embeddings, identifies the most relevant documents or examples from a pre-constructed knowledge base.
\item Generation phase: The retrieved information is incorporated into the prompt, enabling the LLM to generate more informed and contextually appropriate outputs.
\end{itemize}

RAG is particularly effective for tasks requiring domain-specific or up-to-date knowledge, where LLMs alone may struggle due to limitations in their training data~\citep{falke2019ranking,duvsek2020evaluating,brown2020language,liu2023pre,ji2023survey,gao2023retrieval}. By dynamically integrating external knowledge, RAG enhances both the reasoning and contextual understanding of LLMs.
In this work, we employ a RAG approach~\citep{gao2023retrieval,ma2023query} to retrieve examples of annotated S2R entities from a vector database, using them to guide LLMs in accurately extracting and generating actionable operations for bug reproduction.
We adapt RAG to retrieve annotated S2R examples, addressing LLMs' lack of bug report-specific knowledge.

\subsection{Bug Reports and App UI Elements}\label{sec2.4}
Bug reports are essential for documenting unexpected application behaviors. They are typically user-or tester-generated and describe the problem, the conditions under which it occurs, and the S2R. However, many bug reports lack explicit S2R details, which are crucial for replicating and diagnosing the issues. Accurately extracting S2R information from bug reports is challenging but vital for streamlining the bug reproduction and resolution process.

Mobile applications consist of dynamic UI screens, which serve as the visual canvas for implementing app features and are composed of UI components (widgets) such as buttons, text fields, and checkboxes. 
These components allow users to interact with the app and are organized hierarchically within containers (layouts). Each UI screen is represented as a screenshot paired with metadata detailing the hierarchy and attributes of its components. Attributes include the component type (e.g., TextView, Button), label or text, ID, description, visibility, and size.

Interactions with UI components are represented as actions defined by a tuple (target element, action type, value). The target element refers to the UI component (e.g., button or text box), the action type describes the interaction (e.g., ``click'', ``input'', or ``swipe''), and the value is specific to the action (e.g., input text).

In this work, we define S2R entities as including the following components: interactions with UI components, referred to as \textbf{Action}; the target UI elements involved in those interactions, referred to as \textbf{Element}; the specific content associated with the action (e.g., input text), referred to as \textbf{Value}; and the gesture direction when interacting with scrollable elements, referred to as \textbf{Direction}.
Identifying and accurately mapping interactions with these UI components to S2R entities is crucial for automating bug reproduction.

This work focuses on bridging the gap between textual bug reports and actionable UI interactions by extracting S2R entities and correlating them with app UI components, enabling accurate and automated bug reproduction.
This mismatch between unstructured reports and structured UI interactions motivates our approach.

\subsection{Motivation}\label{subsec2.3}

Despite progress in prior research and tool development, fundamental gaps hinder their application to Android crash reproduction. 

Table~\ref{tab:Example Report} provides examples of real-world bug reports, and Table~\ref{tab:s2r_extraction_results} presents a comparison of the corresponding S2R entity extraction results. \textbf{According to the result shown in Table~\ref{tab:s2r_extraction_results}, it is obvious that during the S2R entity extraction stage, prior approaches often fail to extract S2R entities when the action verbs are implicit or absent in the sentence.} Moreover, these methods frequently fail to extract entities when sentences contain multiple components (e.g., ``Secret field'' or ``other required fields'') or actions (e.g., ``enter'' or ``fill'') of the same type, resulting in incomplete or ambiguous extraction results.
\begin{table}[h]
\caption{Examples of Bug Reports}
\label{tab:Example Report}\begin{tabular*}{\textwidth}{@{\extracolsep\fill}p{0.25\textwidth}p{0.70\textwidth}}
\toprule
\textit{\textbf{Example Case}} & \textit{\textbf{Description}} \\
\midrule
\textbf{Example 1} & 
1. Crash application when search. \\
\midrule
\textbf{Example 2} & 
\begin{tabular}[t]{@{}l@{}}
1. On Category screen (either opened from transaction form or from \\ 
\ \ Settings), click on search icon and enter A as search term \\
2. Category A now appears without children \\
3. Tap and hold on category A
\end{tabular} \\
\midrule
\textbf{Example 3} & 
...Enter ``test'' in the ``Secret'' field (and fill other required fields)... \\
\bottomrule
\end{tabular*}
\end{table}

\begin{table}[h]
\caption{S2R entity extraction results comparison}
\label{tab:s2r_extraction_results}
\begin{tabular*}{\textwidth}{@{\extracolsep\fill}p{0.25\textwidth}p{0.70\textwidth}}
\toprule
\textit{\textbf{Method}} & \textit{\textbf{Extraction Results}} \\
\midrule
\multicolumn{2}{c}{\textbf{Example 1 Results}} \\
\hline
ReCDroid & / \\
\midrule
AdbGPT & / \\
\midrule
\textbf{Ours} & \textbf{[Tap]} \textbf{[search]} \\
\midrule
\multicolumn{2}{c}{\textbf{Example 2 Results}} \\
\hline
ReCDroid & 
\begin{tabular}[t]{@{}l@{}}
1. [Tap][screen] \\
2. [Tap][icon] \\
3. [input] \\
4. [click]
\end{tabular} \\
\midrule
AdbGPT & 
\begin{tabular}[t]{@{}l@{}}
1. [Tap][search icon] \\
2. [Input][search term]
\end{tabular} \\
\midrule
\textbf{Ours} & 
\begin{tabular}[t]{@{}l@{}}
\textbf{1.} \textbf{[Tap]} \textbf{[search icon]} \\
\textbf{2.} \textbf{[Input]} \textbf{[search term]} \textbf{[A]} \\
\textbf{3.} \textbf{[Long Tap]} \textbf{[category A]}
\end{tabular} \\
\midrule
\multicolumn{2}{c}{\textbf{Example 3 Results}} \\
\hline
ReCDroid & 
\begin{tabular}[t]{@{}l@{}}
{[Input]}{[field]}{[test]} \\
{[Input]}{[field]}{[test]}
\end{tabular} \\
\midrule
AdbGPT & {[Input]}{[Secret]}{[test]} \\
\midrule
\textbf{Ours} & 
\begin{tabular}[t]{@{}l@{}}
\textbf{[Input]} \textbf{[Secret field]} \textbf{[test]} \\
\textbf{[Input]} \textbf{[other required fields]}
\end{tabular} \\
\bottomrule
\end{tabular*}
\end{table}
During the replay stage, prior methods~\citep{zhao2019recdroid,feng2024prompting} fail to address scenarios requiring multi-page flows or complex functionalities. \textbf{According to the algorithms used in prior approaches, during the exploration process, if a bug report skips the description of a component on the current screen and directly mentions a component on the next screen, the system tends to perform a ``back'' operation to return to the previous screen.} For example, a bug report says ``Select About'', but on the current screen, there is no component labeled ``About''. We need to select any component from this screen first, and then we can see the ``About'' component. When prior methods encounter this situation, they simply choose the ``back'' action, which never leads to the correct screen. This often leads to incorrect looping behavior, ultimately causing the failure of bug reproduction.

To enhance bug reproduction efficiency, we adopt the RAG approach to overcome issues in the S2R entity extraction stage, and leveraging LLMs with GUI-derived app knowledge for guided replay through UTG construction and interface analysis.

\section{Approach}\label{sec3}

\begin{figure}[h]
\centering
\includegraphics[width=130mm]{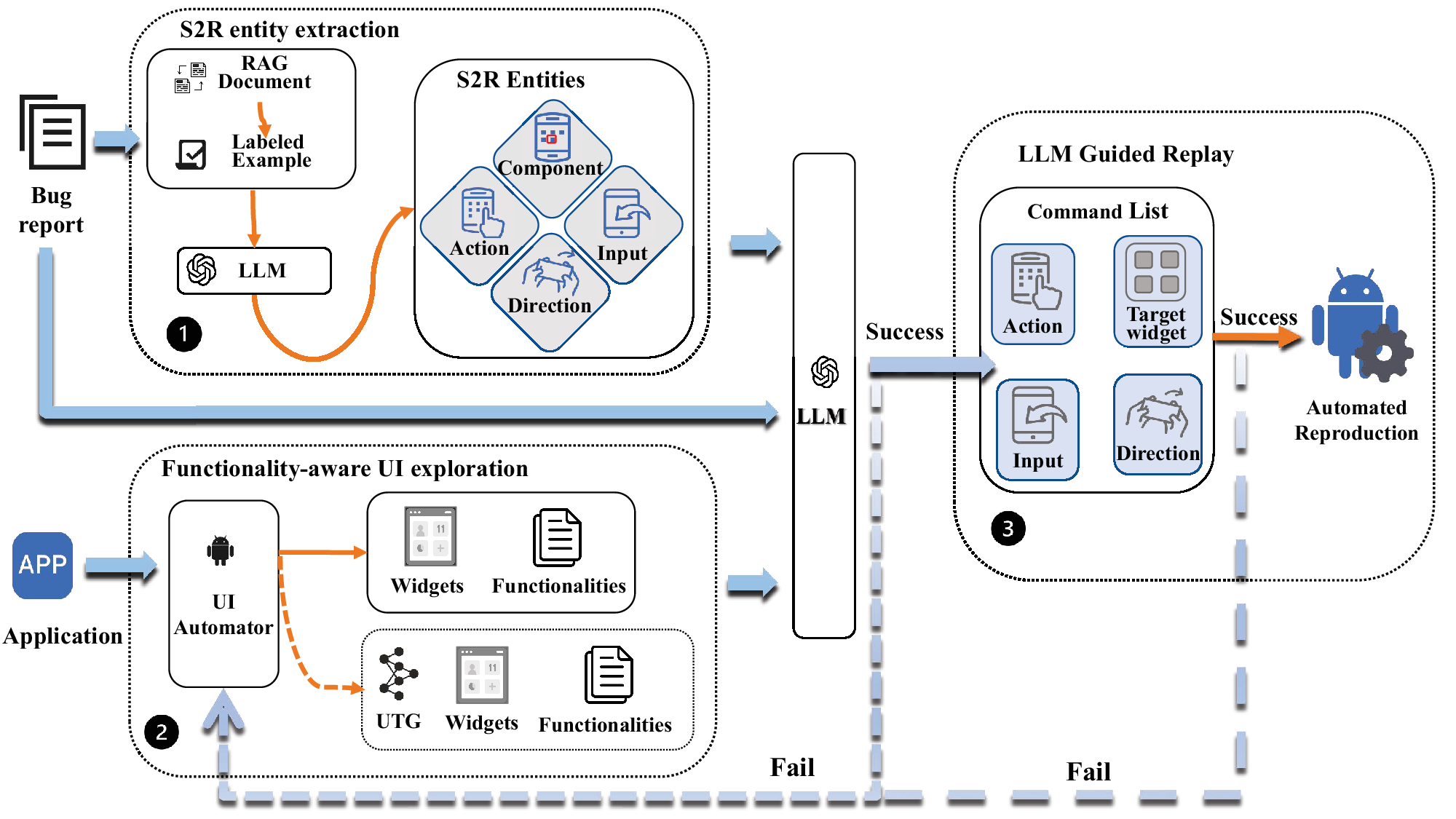}
\caption{The overview of our method.}
\label{fig:1}
\end{figure}

This paper introduces BugRepro, an effective approach to overcoming the challenges posed by ambiguous S2Rs and the intricate nature of modern Android applications. Fig. \ref{fig:1} presents an overview of BugRepro. In the following sections, we detail the two main components of BugRepro: RAG-enhanced S2R entity extraction (§~\ref{sec3.1}) and exploration-based replay (§~\ref{sec3.2}).

Our method is structured into two distinct phases and comprises three core components. \textbf{The first phase is RAG-enhanced S2R entity extraction,} we employ RAG to enhance the accuracy of S2R entity extraction (§~\ref{sec3.1.2}). When a new bug report is submitted, it is first segmented into individual sentences. For each sentence, the system automatically retrieves semantically similar sentences and their corresponding labels (which include all associated S2R entities) from the RAG database---an example is shown in Table \ref{tab:retrieval_example}. Then, the original bug report along with the retrieved labeled examples is fed into LLM, which is guided by the examples to analyze and extract the S2R entities corresponding to the new bug report. After extracting the S2R entities, we leverage LLMs to interact with UI components based on these entities and corresponding bug reports to replay the bug. At the same time, based on existing work and our observations, we found that even after applying RAG, challenges remain in extracting components, which may lead to difficulties in correctly matching and operating the corresponding components during the replay process. Therefore, we need to provide the LLM with more UI-related knowledge. \textbf{The second phase is Exploration-based Replay,} which aims to provide app-related knowledge to LLMs, enabling them to gain insights into the app, understand its usage, and make informed decisions. \textbf{It consists of two primary components: UI exploration based on functionality and LLM guided replay, which continuously interact with each other throughout the process;} \textbf{The functionality-aware UI exploration component} is responsible for exploring the UI, and all information will be transmitted to LLM. Note that since UI screens are visual but can not be processed by LLMs directly, we encode the UI screens into text format for LLMs processing following existing work~\citep{feng2024prompting}. \textbf{The LLM guided replay component} receives the encoded UI information from the functionality-aware UI exploration module, synthesizes it with inputs from the RAG-enhanced S2R entity extraction module and the original bug report, and then generates guidance for the next interaction step based on this comprehensive context. 

\textbf{After executing each step, Bugrepro checks whether the bug reproduction has succeeded and provides feedback to the LLM. If reproduction fails, it triggers the functionality-aware UI exploration process, allowing the acquisition of additional domain-specific knowledge to enhance the LLM’s decision-making.} During the LLM’s decision-making process, when LLM faces difficulties determining the next step (§~\ref{sec3.2.1}), BugRepro augments its decision-making by integrating app-specific knowledge, which encompasses both the UI exploration and the synthesis of app-related information. This process involves analyzing the app’s UI elements and interactions to construct the Synthesized Functionality Table (Table \ref{tab:synthesized-functionality}), which consists of three components: \texttt{<Synthesized functionality, UI states, UI elements>}. Once equipped with more comprehensive domain-specific knowledge, the LLM is able to make more accurate and effective decisions in response to the current situation.

\subsection{RAG-enhanced S2R entity extraction}
\label{sec3.1}

As discussed in Section~\ref{sec1}, LLMs struggle with accurate S2R entity extraction due to their lack of bug-report-specific knowledge. Without the background context or established best practices for extracting these entities, LLMs often produce inaccurate results. 
To overcome this, BugRepro enhances S2R entity extraction by integrating LLMs with relevant, bug-report-specific knowledge using a RAG approach.
In particular, BugRepro incorporates a lightweight RAG component that retrieves task-relevant information (e.g., similar bug reports and their corresponding labeled S2R entities). This retrieved information is then incorporated into the LLM’s prompt, equipping the model with both bug-report-specific and task-oriented knowledge, which significantly improves its contextual understanding and ensures more accurate extraction of S2R entities.
In the following, we first describe how we construct the RAG database and then explain how it enhances the S2R entity extraction process.

\subsubsection{Constructing the RAG database}
The Basic RAG approach stores all documents in a single database, reflecting the most common cases of RAG-based LLMs~\citep{lewis2020retrieval,ma2023query,gao2023retrieval}. Therefore, we first construct such a database for the RAG process.

\textbf{Data Collection}.
We begin by crawling a large collection of bug reports from various web resources referenced in prior research~\citep{zhao2019recdroid,su2021benchmarking,cooper2021takes,huang2023context}. 
These reports are then embedded to construct a vector database for the RAG process. 
Specifically, we use the all-MiniLM-L12-v2 model~\citep{sentence-transformers-all-MiniLM-L6-v2,galli2024performance,aperdannier2024systematic,park2024adaptive}, which is known for its lightweight architecture, computational efficiency, and high accuracy. 
It is worth noting that the choice of embedding model is orthogonal to our approach, and it can easily be swapped for other embedding models.

\textbf{Labeling S2R Entities}.
Although we obtain a set of bug reports in the previous step, they are not annotated with the corresponding S2R entities.
Following existing work~\citep{feng2024prompting, wang2024feedback}, we employ four developers with 6--10 years of experience in both software testing and development to label the bug reports.

Before starting the labeling process, we provide the developers with a detailed project overview. Through a lottery process, three developers are selected to label the reports, while the fourth developer is tasked with consolidating their results.
To label those bug reports, the developers first carefully break down each bug report into individual sentences. 
The developers then carefully analysis each sentence to extract its corresponding S2R entities, which consist of four distinct components, as defined by our predefined standards: action type, target component, input values, and scroll direction.

\textbf{Cross-Validation and Consistency Assurance}.
Once the initial labeling is complete, following existing work~\citep{feng2024prompting, wang2024feedback}, the fourth developer assumes a crucial role in ensuring the consistency and accuracy of the labeled data.  We follow the dataset construction approach used in AdbGPT, which involves hiring two data annotators. When their annotations differ, a third person is brought in to make the final decision. Specifically, the fourth developer systematically compares the results produced by the first three developers, focusing on resolving discrepancies in how identical sentences are interpreted. Leveraging extensive expertise, the fourth developer unifies the varying opinions, ensuring the final labeled dataset is scientifically consistent and reliable.
\subsubsection{RAG-enhanced S2R entity extraction}\label{sec3.1.2}
Given a bug report \textbf{\boldmath$B$} consisting of \textbf{\boldmath$N$} sentences, denoted as \textbf{\boldmath${s_1, s_2, \dots, s_n}$}, BugRepro first divides the report into individual sentences. 
Next, BugRepro uses the same embedding model employed during the database construction (i.e., all-MiniLM-
L12-v2 model) to embed each sentence. 
For each sentence \textbf{\boldmath$s_i$ (\( i = 1, 2, \dots, n \))}, BugRepro calculates the cosine similarity between the embedding of \( s_i \) and each sentence in the RAG database. The most similar sentence, along with its annotated S2R entity, is then retrieved and used as a reference for extracting the corresponding S2R entity for \textbf{\boldmath\( s_i \)}.

\begin{table}[h]
\caption{An example of retrieved results.}
\label{tab:retrieval_example}
\begin{tabular*}{\textwidth}{@{\extracolsep\fill}p{0.45\textwidth}|p{0.45\textwidth}}
\toprule
\textit{\textbf{Sentence in current Bug Report}} & \textit{\textbf{Retrieved sentence with S2R entity}} \\
\midrule
Long hold on any video and press add to playlist. (Alternatively, select the add to playlist option under any video when watching). & 
Sentence: Click on the add to option and select playlist. \newline
The entities extracted from the sentence are: \newline
1. [Tap] [add to option] \newline
2. [Tap] [playlist] \\
\bottomrule
\end{tabular*}
\footnotetext{Note: This table shows a example of a bug report sentence and a matched sentence retrieved from the RAG database with extracted S2R entities.}
\end{table}

Table~\ref{tab:retrieval_example} provides an example of a retrieval result. Given a sentence from a bug report (e.g., ``Long hold on any video and press add to playlist. Alternatively, select the add to playlist option under any video when watching.''), BugRepro retrieves the most similar sentence from the database (e.g., ``Click on the add to option and select playlist.'') and then, the corresponding S2R entities. 
Incorporating this retrieved example into the prompt assists the LLM in accurately extracting the S2R entity for the given sentence.

\textbf{Prompt Construction}.
After retrieving relevant examples for each sentence in the bug report, BugRepro constructs a prompt to instruct the LLM to extract S2R entities.
Following prior work~\citep{feng2024prompting}, we begin by identifying all available actions and action primitives within the S2R entities to guide the LLM. 
Specifically, we define seven standard actions: \textbf{[tap, input, scroll, rotate, delete, double tap, long tap]}, which we refer to as \textbf{available actions}. The context of each action may vary, and thus each action requires a set of primitives to accurately represent its entities. 
For instance, the \textbf{Tap} action requires a target component to interact with, such as a button in the GUI. This can be represented as \textbf{[Tap] [Component]}. Similarly, for the \textbf{Double-tap} and \textbf{Long-tap} actions, we use analogous linguistic primitives. 
The \textbf{Scroll} action involves specifying the direction of the scroll, such as upward or downward, which we formulate as \textbf{[Scroll] [Direction]}. Likewise, the \textit{Rotate} action is expressed using a similar structure, i.e., \textbf{[Rotate] [Direction]}.
The \textbf{Input} action corresponds to entering a specific value into a text field and is thus formulated as \textbf{[Input] [Component] [Value]}. Similarly, the \textbf{Delete} action is represented using the primitive \textbf{[Delete] [Component]}, and other similar actions follow the same pattern.

After defining the available actions and their corresponding primitives, BugRepro incorporates the retrieved examples into the prompt to further assist the LLM. 
Specifically, BugRepro includes each retrieved sentence along with its corresponding annotated S2R entity, formatted as in the following examples for S2R entity extraction. The sentence is ``Click on the add to option and select playlist.'', the extracted S2R entities are: 1. [Tap] [add to option], 2. [Tap] [playlist].

Table~\ref{tab:prompt_design} shows an example of the constructed prompt. The aforementioned information is aggregated with the content of \textbf{\boldmath{$b$}} to form the complete prompt. The final structure of the prompt sent to the LLM includes:
\textbf{⟨Available Actions⟩ + ⟨Action Primitives⟩ + ⟨Retrieved Example Sentences with S2R Entities⟩ + ⟨Current Bug Report⟩}.

\begin{table}[h]
\caption{An example of designing prompts for S2R entity extraction.}
\label{tab:prompt_design}
\begin{tabular*}{\textwidth}{@{\extracolsep\fill}p{0.25\textwidth}p{0.70\textwidth}}
\toprule
\textit{\textbf{Prompt Framework}} & \textit{\textbf{Implementation}} \\
\midrule
\textbf{Available\_Actions} & 
[tap(click), input(set\_text), scroll, swipe, rotate, delete, double tap(click), long tap(click), restart, back]. 
Generate input when none is given. \\
\midrule
\textbf{Action\_Primitive} & 
[Tap] [Component], [Scroll] [Direction], [Input] [Component] [Value], [Rotate] [Component], [Delete] [Component] [Value], [Double-tap] [Component], [Long-tap] [Component].  
The actions you identify should be in the available actions. \\
\midrule
\textbf{Retrieval\_Prompt} & 
Here are some examples for S2R entity extraction.  
The sentence is ``Click on the add to option and select playlist.'', the extracted S2R entities are:  
1. [Tap] [add to option]  
2. [Tap] [playlist].
The sentence is ``Rotate your phone. (Enable auto-rotate first).'', the extracted S2R entity is:  
1. [Rotate]. \\
\midrule
\textbf{Current\_Bug\_Report} & 
Here are the sentences in current bug report:  
1. Long hold on any video and press add to playlist.  
(Alternatively, select the add to playlist option under any video when watching).  
2. Rotate the screen while auto-rotate feature is on. \\
\bottomrule
\end{tabular*}
\end{table}

\subsection{Exploration-based Replay}
\label{sec3.2}

\subsubsection{UTG-Driven Functional Exploration}
\label{sec3.2.1}
UI exploration is triggered when the LLM encounters uncertainty regarding which UI component to interact with. 
Therefore, BugRepro begins the exploration by analyzing the UTG starting from the page where the difficulty arises, denoted as \textbf{\boldmath$p$}. 
The UTG, created by the UI  Automator, provides essential information about the app, including the relationships between UI states and the presence of various UI elements on each screen. By examining the functionalities of these elements, we can uncover the tasks that can be performed within the app and identify the UI components necessary to execute them. Therefore, BugRepro parses the UI states and elements in the UTG, querying LLMs to extract their respective functions.

In this context, the UTG is modeled as a directed graph, where nodes represent UI states and edges represent actions, both recorded by the random Explorer. For each UI state \textbf{\boldmath\( U_i \)}, BugRepro queries the LLM to summarize the functionalities of all UI elements \textbf{\boldmath\( e_i \)} within that state. 
Notably, BugRepro only extracts the functionality of an element from the UI state closest to the initial UI if it appears in multiple UI states.

Table~\ref{tab:synthesized-functionality} shows the construction of the synthesized functionality table. Upon traversing the entire UTG, BugRepro compiles a synthesized functionality table in the app-related knowledge, which contains $\boldsymbol{n}$ entries, where $\boldsymbol{n}$ is the total number of UI elements in the page $\boldsymbol{p}$. Each entry corresponds to a UI element $\boldsymbol{e_i}$ and includes three components: \texttt{<Synthesized functionality, UI states, UI elements>}. 

\begin{table}[hbtp]
\caption{A portion of the Synthesized Functionality Table}
\label{tab:synthesized-functionality}
\footnotesize
\setlength{\tabcolsep}{3.5pt} 
\begin{tabular}{|l|c|l|c|p{0.33\textwidth}|}
\hline
\multicolumn{2}{|c|}{\textit{\textbf{Problem Page}}} & \multicolumn{3}{c|}{\textit{\textbf{Other UI States}}} \\
\hline
\bfseries\begin{tabular}{@{}l@{}}Current \\ Activity\end{tabular} & 
\bfseries Elements & 
\bfseries\begin{tabular}{@{}l@{}}Current \\ Activity\end{tabular} & 
\bfseries Elements & 
\bfseries Synthesized Functionality \\
\hline
\begin{tabular}{@{}l@{}}MainFlash\\Activity\end{tabular} & 
INFO & 
\begin{tabular}{@{}l@{}}Welcome\\Activity\end{tabular} & 
\begin{tabular}{@{}l@{}}TextView/Static Text,\\ Button,\\ Status Bar,\\ Navigation Bar\end{tabular} & 
\begin{tabular}{@{}p{0.33\textwidth}@{}}
This is the ``About LibreNews'' screen in the LibreNews app. It provides information about the app's features and purpose, including that it offers ad-free, decentralized, and secure breaking news notifications. The screen explains that LibreNews is open source, doesn't track users, and primarily interacts with users through notifications (about 3 per day with default settings). The main actionable element is a ``GO TO LIBRENEWS'' button at the bottom of the screen.
\end{tabular} \\
\hline
\begin{tabular}{@{}l@{}}MainFlash\\Activity\end{tabular} & 
REFRESH & 
\begin{tabular}{@{}l@{}}Welcome\\Activity\end{tabular} & 
\begin{tabular}{@{}l@{}}TextView/Static Text,\\ Button,\\ Status Bar,\\ Navigation Bar\end{tabular} & 
\begin{tabular}{@{}p{0.33\textwidth}@{}}
This screen displays the ``About LibreNews'' page within the LibreNews application. It provides information about the app's features including being ad-free, decentralized, secure breaking news notifications, and its open-source nature. The page contains several text sections explaining the purpose and benefits of LibreNews, and a prominent ``GO TO LIBRENEWS'' button at the bottom that is clickable. The system UI elements visible include the status bar (with time, wifi, mobile signal, and battery indicators) and the navigation bar (with back, home, and recent apps buttons).
\end{tabular} \\
\hline
\end{tabular}
\footnotetext{\footnotesize Note: This table presents a subset of the Synthesized Functionality Table. Here, the \textbf{Problem Page} denotes the UI page where the issue was initially observed. The corresponding \textbf{Current Activity} specifies the Android Activity associated with that page. \textbf{Elements} refer to individual UI components present on the page. Through Functionality-aware UI exploration, we identify the subsequent \textbf{UI states} triggered by interacting with these elements from the Problem Page, and extract their corresponding \textbf{New Activity, Elements, and the Synthesized Functionality as inferred by the LLM.}}
\end{table}

The ``Synthesized functionality'' represents the task associated with $\boldsymbol{e_i}$, as summarized by the LLM, which can be accomplished by interacting with this element. ``UI elements'' refers to the elements that are clicked to navigate from the initial UI state to $\boldsymbol{U_i}$, and ``UI states'' represent the sequence of UI states traversed during this process. This table provides the LLM with the necessary information to determine the actions, facilitating more efficient planning.

In addition to the synthesized functionality table, the app-related knowledge also includes a UI function table, which summarizes the functionality of each UI state in the UTG. This information is derived by querying the LLM to describe the function of each UI state in the graph.

\subsubsection{Exploration-enhanced Prompting}

Typically, to match the extracted S2R entities to a sequence of GUI events for bug reproduction, a common solution is to use lexical computation to match the extracted components against the displayed text of the UI components on the current screen~\citep{fazzini2018automatically,zhao2019recdroid,feng2024prompting}. However, this approach can be inaccurate due to the absence of textual descriptions or vague bug reports~\citep{white2019improving,cooper2021takes}. To overcome this limitation, we utilize LLMs to generate dynamic guidance on the GUI screen, enabling automatic reproduction of the steps.

Specifically, we begin by providing LLMs with instructions about the bug reproduction task. These instructions serve as a foundational guide for the LLM, outlining the objective, workflow, and detailed steps for reproducing Android bugs.
Next, we supply the LLM with the current bug report, the required S2R entities, and the encoded text of the current UI screen, allowing the LLM to determine which UI component to interact with and what action to perform. Specifically, our prompt consists of the following components: (1) the bug report text, (2) the extracted S2R entity, and (3) the encoded text of the current UI screen.

However, when the LLM struggles to make the correct decision---such as when it cannot determine which UI component to interact with, or when exploration based on the LLM's guidance fails to reproduce the bug and the LLM does not update its decision after revisiting the page. In such cases, we provide additional app-related knowledge obtained from the previous exploration to assist the LLM. Specifically, given a problematic page, we supply the LLM with the functionality of all UI components on that problematic page and the functionalities of the new UI screens they can navigate to. This additional information enables the LLM to decide which UI component to interact with and what action to take. At this stage, the prompt includes the following components: (1) the bug report text, (2) the extracted S2R entity, (3) the encoded text of the current UI screen, and (4) the app-related knowledge, which includes the synthesized functionalities (§~\ref{sec3.2.1}) of each UI component within current UI page, and the UI function table, which includes a mapping of functionalities associated with UI pages accessible through specific components on the current page.

\subsubsection{Interpreting the response}\label{sec3.2.3}
Due to the inconsistent output formats of LLMs in prior work, it has been challenging to bridge the gap between LLM-generated responses and structured executable code. To address this issue, we adopt the LangChain framework to enforce a standardized output structure from the LLM.

Table~\ref{tab:json-table} illustrates the form a response may take. During each iteration of the bug reproduction process, the system relies on the LLM’s response to execute actions on the current page. The response is consistently represented as a JSON-formatted array, regardless of whether it contains a single action primitive or a sequence of multiple actions.  Each action primitive includes essential fields such as the action type, the target UI component (represented as \textbf{feature}), and, when applicable, additional parameters like input text, direction, or duration.

\begin{table}[htbp]
\caption{Example of a JSON-formatted action sequence generated by LLM}
\label{tab:json-table}
\begin{tabular}{|l|l|}
\hline
Single Action Primitive &
  \begin{tabular}[c]{@{}l@{}}{[}\\     \{\\         ``action'': ``click'',\\         ``feature'': ``REFRESH''\\     \}\\ {]}\end{tabular} \\ \hline
Sequences of Action Primitives &
  \begin{tabular}[c]{@{}l@{}}{[}\\     \{\\         ``action'': ``set\_text'',\\         ``feature'': ``https://librenews.io/api'',\\         ``input\_text'': ``xxyyzz''\\     \},\\     \{\\         ``action'': ``click'',\\         ``feature'': ``OK''\\     \}\\ {]}\end{tabular} \\ \hline
\end{tabular}
\end{table}

These generated actions are parsed using regular expressions based on predefined patterns. After interpreting the generated actions, the system executes them and provides immediate feedback to the LLM regarding the execution status—indicating whether the actions were successfully carried out. This feedback enables the LLM to assess the current UI state and decide whether to proceed to the next step or revise its response in case of a failed execution. The reproduction loop continues iteratively until either the bug is successfully reproduced or the process terminates due to exceeding the allotted time budget.

After interpreting the generated actions, BugRepro executes them and provides feedback to the LLM regarding the execution status, indicating whether the actions were successfully carried out. This feedback allows the LLM to assess the current status and decide whether it is appropriate to proceed to the next step or if the response needs to be reformulated due to a failed execution.
The reproduction process continues iteratively until either a successful reproduction is achieved or the bug fails to be reproduced within the allocated time budget.

\section{Experimental Setup}\label{sec4}
To evaluate BugRepro, we consider four key research questions:
\begin{itemize}
    \item \textbf{RQ1: How accurate is BugRepro in extracting S2R entities?}
    This investigates the performance of our S2R entity extraction approach, emphasizing the impact of RAG and few-shot examples. Comparisons are made with state-of-the-art baselines to highlight BugRepro's advantages. For more details, please refer to Section \ref{sec5.1}.

\item \textbf{RQ2: How effective and efficient is BugRepro in reproducing bug reports?}
This examines the ability of our approach to reproduce crashes from bug reports within a predefined time limit, emphasizing both success rate and efficiency. For more details, please refer to Section \ref{sec5.2}.

\item \textbf{RQ3: How do individual components impact performance when removed?}
Through ablation studies, this explores the contribution of each innovation in our approach to S2R entity extraction and bug reproduction. For more details, please refer to Section \ref{sec5.3}.

\item \textbf{RQ4: How does the robustness of each method vary across different language models?}
This investigates the effect of using different LLMs (e.g., GPT-4, DeepSeek) on the performance of our approach. For more details, please refer to Section \ref{sec5.4}.
\end{itemize}

\subsection{Datasets}
To construct our dataset, we follow established practices for collecting real-world bug reports for reproduction studies~\citep{zhao2019recdroid,su2021benchmarking,huang2023context,cooper2021takes}. Specifically, to minimize potential bias, we source bug reports from four well-known open-source datasets: (i) the evaluation dataset of ReCDroid~\citep{zhao2019recdroid}; (ii) the evaluation dataset of ScopeDroid~\citep{huang2023context}; (iii) the empirical study dataset from AndroR2~\citep{wendland2021andror2}; and (iv) another empirical study on Android bug report reproduction~\citep{su2021benchmarking}.
Given the overlap across these datasets, we first eliminate duplicate entries, ensuring that no bug reports from the same issue repository were counted multiple times. While users often include visual elements such as screenshots or videos in their bug reports~\citep{bernal2020translating,kuramoto2022visual,cooper2021takes,fang2021read,feng2022gifdroid}, our focus in this work is on textual information, specifically natural language descriptions of S2Rs. Consequently, we manually review the bug reports and retain only those containing textual S2Rs.
After these filtering steps, our final experimental dataset comprises 151 bug reports. This curated collection serves as a reliable foundation for evaluating our approach.

\subsection{Implementation and Environment}
We implement BugRepro in Python, and we leverage all-MiniLM-
L12-v2 model as the retrieval model used in the RAG component, which is based on the Dense Passage Retrieval (DPR)~\citep{karpukhin2020dense} architecture and uses a bi-encoder to encode the queries and documents into dense vectors.
We utilize the DeepSeek-V3 model as the underlying LLM in BugRepro, accessed via its API service~\citep{deepseek2024}. To address potential verbosity in the LLM’s output—such as repeated questions or chain-of-thought reasoning—we adopt a filtering mechanism that removes non-JSON content, retaining only valid JSON segments enclosed in square brackets.
For interacting with UI widgets on the device, we leverage UI Automator2~\citep{uiautomator2} as the execution engine. 
We set the UTG exploration depth to 1 and the time budget to 5 minutes to balance effectiveness and efficiency.
All experiments are performed on a workstation running Windows 11 Pro 64-bit, equipped with a 13th Gen Intel(R) Core(TM) i7-13700KF CPU (24 cores, base frequency ~3.4GHz) and an NVIDIA GeForce RTX 4090 GPU.

\subsection{Baselines}
We select two state-of-the-art methods widely recognized for android bug reproduction to compare against BugRepro: 1) \textbf{ReCDroid}~\citep{zhao2019recdroid}: This is a traditional state-of-the-art method that analyzes dependencies among words and phrases from hundreds of bug reports and uses 22 predefined grammar patterns to extract S2R entities. For example, a noun phrase (NP) followed by a ``click'' action is interpreted as the target component. We adopt their released repository for evaluation. Although ReCDroid+~\citep{zhao2019recdroid} is an extended version of ReCDroid, it primarily focuses on scraping bug reports from issue tracking websites, which falls outside the scope of this study. We use ReCDroid in this work for simplicity.  
2) \textbf{AdbGPT}~\citep{feng2024prompting}: This is a state-of-the-art LLM-based method leveraging few-shot learning and chain-of-thought reasoning to extract S2R entities from bug reports effectively.  

To understand the contributions of individual components in BugRepro, we develop two ablation variants of BugRepro: 1) \textbf{BugRepro$_{nor}${}}: This variant omits the retrieval model and relies solely on the LLM-based generation model to produce actionable operations.
2) \textbf{BugRepro$_{nou}${}}: This variant excludes the UI exploration process and retrieves S2R entities directly from the database using the retrieval model alone. 

To evaluate the generalization of BugRepro, we replace the default LLM (DeepSeek-V3) with GPT-4, naming this variant \textbf{BugRepro$_{gpt}${}}. 
This allows us to assess how different LLMs affect the robustness and performance of BugRepro.

\subsection{Metrics}
Following existing work~\citep{feng2024prompting}, we measure the effectiveness of S2R entity extraction using the \textbf{accuracy} metric. An extracted S2R is deemed accurate if all of the following match the ground truth, including steps (i.e., step ordering, sub-step separation) and entities (i.e., the action types of each step, the possible target components if existed, the possible input values, and the possible scroll directions). 
A higher accuracy score reflects an approach’s superior ability to correctly interpret and extract the steps-to-reproduce (S2R) entity from bug reports.
To evaluate the effectiveness of reproducing bugs, we use the \textbf{number of successful reproductions (NSR)}.
A higher NSR indicates better performance in replicating the S2Rs on GUIs to successfully trigger the reported bugs.
To assess the efficiency of each technique, we measure the \textbf{average time} taken for successful reproductions.
Less time indicates greater efficiency in reproducing the bugs, highlighting the method's practical applicability in real-world scenarios.

\section{Results}\label{sec5}
\subsection{RQ1: How accurate is BugRepro in extracting S2R entities?}\label{sec5.1}

\begin{table}[t]
\centering
\caption{Accuracy Comparison on S2R Entity Extraction}\label{tab:s2r_performance}
\begin{tabular*}{\columnwidth}{@{\extracolsep\fill}lrrrrr}
\toprule
Method   & Step & Action & Component & Input & Direction \\
\midrule
Recdroid & 63.02\% & 45.50\% & 4.98\% & 52.00\% & 0.00\% \\
AdbGPT   & 76.16\% & 61.92\% & 28.77\% & 62.08\% & 73.12\% \\
\textbf{Ours} & \textbf{87.85\%} & \textbf{69.49\%} & \textbf{33.87\%} & \textbf{70.93\%} & \textbf{82.91\%} \\
\bottomrule
\end{tabular*}
\footnotetext{Note: This table shows the accuracy comparison of different methods for S2R entity extraction across various categories.}
\end{table}

To answer RQ1, we evaluate the accuracy of BugRepro in extracting S2R entities compared to the state-of-the-art baselines.
Table~\ref{tab:s2r_performance} presents the comparison results of BugRepro, AdbGPT, and ReCDroid across five dimensions: step extraction, action types, target components, input values, and action direction.

In the table, BugRepro outperforms both ReCDroid and AdbGPT in all dimensions, with 7.57\% to 11.69\% higher accuracy than AdbGPT. For instance, BugRepro records 69.49\% accuracy in extracting action types, compared to 61.92\% by AdbGPT and 45.50\% by ReCDroid. Similarly, in identifying target components, BugRepro achieves 33.87\%, outperforming AdbGPT (28.77\%) and ReCDroid (4.98\%). This trend persists across input values and scroll directions, underscoring BugRepro's consistent advantage.

Through our analysis, there are three main reasons that BugRepro outperforms AdbGPT and ReCDroid:
\begin{itemize}
    \item Enhanced Contextual Understanding: 
    These improvements demonstrate BugRepro's ability to overcome limitations inherent in traditional grammar-based methods (ReCDroid) and static few-shot learning approaches (AdbGPT). The RAG component is pivotal in achieving this performance, enabling the retrieval of task-relevant examples to provide richer context for the LLM. The use of RAG allows BugRepro to bridge the gap between generic LLM capabilities and the domain-specific needs of S2R extraction. 
    \item Precision in Multi-Step Scenarios: For bug reports requiring multi-step actions, BugRepro excels in maintaining the correct sequence and sub-step separation, which are critical for accurate S2R entity extraction. In contrast, AdbGPT often struggles with multi-step reasoning due to its reliance on few-shot learning without dynamic retrieval.
    \item Robustness to Linguistic Variations: BugRepro demonstrates a high degree of adaptability to different writing styles, including reports with incomplete or colloquial descriptions. This flexibility is attributed to the diverse examples provided by RAG, which act as a bridge to align the model’s predictions with the intended semantics.
\end{itemize}

\fbox{%
    \parbox{0.95\linewidth}{%
        \textbf{RQ1 summary:} Integrating RAG into our approach clearly demonstrates that BugRepro outperforms baseline methods in S2R extraction, achieving accuracy improvements of 7.57\% to 11.69\% over AdbGPT (state-of-the-art baselines) across all dimensions. This fully confirms that RAG not only provides contextualized examples, enabling BugRepro to effectively handle ambiguous or complex bug reports, but also enhances contextual understanding, precision in multi-step scenarios, and robustness to linguistic variations.
    }%
}

\subsection{RQ2: How effective and efficient is BugRepro in reproducing bug reports?}\label{sec5.2}
To answer RQ2, we evaluate the effective: \textbf{NSR} and efficiency of BugRepro in reproducing bugs compared to baselines.
Table~\ref{tab:NSR_efficiency} provides the comparison results. From the table, BugRepro achieves an NSR of 96/151, significantly higher than AdbGPT 55/151 and ReCDroid 38/151.

\begin{table}[h]
\caption{NSR and Extract Time in Reproducing Crashes}\label{tab:NSR_efficiency}
\begin{tabular*}{\textwidth}{@{\extracolsep\fill}lcc}
\toprule%
Method & NSR & Average Time\\
\midrule
ReCDroid & 38 & 534.9s \\
AdbGPT  & 55 & 125.6s \\
\textbf{Ours}    & \textbf{96} & \textbf{124.7s} \\
\botrule
\end{tabular*}
\end{table}

BugRepro's advantage is particularly evident in scenarios involving complex app workflows or ambiguous S2Rs. 
One common limitation in existing methods is their reliance solely on the content of bug reports for S2R entity extraction, often neglecting the domain-specific context required for precise interpretation. This can result in inaccuracies during the extraction phase. By incorporating RAG, BugRepro enriches the LLM’s understanding with task-relevant knowledge, effectively addressing these challenges and improving accuracy.

Moreover, existing approaches, such as AdbGPT and ReCDroid, follow strictly sequential steps within the current screen, focusing on locating specific UI widgets. These methods rarely consider transitioning to a different screen unless explicitly guided by predefined heuristics, leading to limited exploration capabilities. In contrast, BugRepro integrates UTG to provide a broader understanding of the app’s structure. This enables it to navigate across multiple screens dynamically and equips the LLM with synthesized app-specific knowledge from this process, significantly reducing redundant exploration and improving precision.

For instance, in a case requiring navigation through three app screens before encountering a bug, BugRepro effectively utilizes its synthesized app knowledge to streamline the reproduction process. On the other hand, AdbGPT and ReCDroid frequently revisit previously explored states or fail to identify the correct sequence, leading to lower success rates.

The integration of domain-specific knowledge with comprehensive UI contextual analysis enables BugRepro to exhibit superior performance in crash reproduction scenarios. The empirical evaluation demonstrates that BugRepro successfully reproduces 96 crash cases, representing a 152.63\% improvement over ReCDroid (38) and a 74.55\% improvement over AdbGPT (55) in reproduction capability. In terms of temporal efficiency, BugRepro completes reproduction in 124.7 seconds on average, which is 4.29 times faster than ReCDroid (534.9s) and marginally 0.7\% faster than AdbGPT (125.6s). This performance makes BugRepro an optimal solution for software engineers engaged in debugging and quality assurance processes, where both comprehensive crash reproduction and time efficiency are critical factors.

\fbox{%
    \parbox{0.95\linewidth}{%
        \textbf{RQ2 summary:} BugRepro significantly outperforms baselines in bug reproduction success rate and efficiency, achieving the NSR of 96 and an average reproduction time of 124.7 seconds per bug. These results highlight the importance of integrating RAG and UTGs, which enable BugRepro to resolve ambiguities, navigate complex app workflows, and minimize redundant exploration. BugRepro offers a practical and efficient solution for real-world debugging tasks.
    }%
}

\subsection{RQ3: How do individual components impact performance when removed?}\label{sec5.3}
To answer RQ3, we conduct ablation studies by removing the RAG component and the UI exploration mechanism independently to assess the contribution of BugRepro’s core components. This analysis quantifies the individual impact of these components on BugRepro’s overall performance in bug reproduction tasks.
Table~\ref{tab:ablation} presents the results, showing a significant decline in number of successful reproduction (NSR) when either component is excluded. The findings highlight the complementary roles of RAG and UI exploration in enabling BugRepro to extract accurate S2Rs and effectively reproduce bugs.

\begin{table}[h]
\caption{Comparison between BugRepro and its variants}\label{tab:ablation}
\begin{tabular*}{\textwidth}{@{\extracolsep\fill}lccc}
\toprule%
& \multicolumn{1}{@{}c@{}}{NSR} & \multicolumn{1}{@{}c@{}}{Average Time} & \multicolumn{1}{@{}c@{}}{Average Response Time\footnotemark[1]} \\
\midrule
BugRepro$_{nor}$  & 80 & 75.2s & 5.99s \\
BugRepro$_{nou}$  & 76 & 62.7s & 5.64s \\
\textbf{Ours}    & \textbf{96} & \textbf{124.7s} & \textbf{9.37s} \\
\botrule
\end{tabular*}

\footnotetext[1]{Average Response Time represents the total time spent by the LLM to generate responses during the crash reproduction process.}
\end{table}

Without RAG, BugRepro$_{nor}$'s NSR drops to 80/151, compared to the BugRepro's 96/151. While the average reproduction time decreases significantly to 75.2 seconds from 124.7 seconds, this efficiency comes at the cost of accuracy, with 15 fewer successful reproductions.

The absence of RAG removes the retrieval of task-relevant examples, which are critical for enhancing the LLM's contextual understanding during S2R entity extraction. This deficiency becomes particularly evident in linguistically complex or ambiguous instructions. For example, phrases such as ``swipe left in the settings'' are frequently misinterpreted as [Swipe] [Left].

However, with the support of retrieved examples, such as ``press the settings icon and rotate the phone,'' where the extracted S2R entities are [Tap] [Settings] and [Rotate], the model gains the contextual understanding needed to correctly interpret the instruction. Consequently, it is more likely to extract the intended S2R sequence as [Tap] [Settings] followed by [Swipe] [Left].

Similarly, instructions involving conditional steps or implicit transitions are less likely to be correctly parsed, resulting in incomplete or incorrect S2Rs and subsequent reproduction failures. The faster LLM response time of 5.99s compared to 9.37s further confirms that without RAG, the model performs less complex reasoning but produces less accurate results.

The removal of UI exploration also significantly impacts performance, with BugRepro$_{\text{nou}}$ reducing the NSR to 76/151 and decreasing the average reproduction time to 62.7 seconds. This highlights the critical role of app-specific knowledge in achieving reproduction accuracy, even though the process may execute more quickly without UI exploration overhead.

UI exploration enables BugRepro to construct and leverage UTGs, which map the dynamic states and interactions within an app. Without this mechanism, BugRepro is limited to interpreting S2Rs in isolation, often failing to navigate multi-screen workflows. For instance, reproducing a bug that requires traversing through nested menus before triggering the issue becomes significantly more challenging without a transition graph to guide the process. This limitation leads to a higher likelihood of failure as the system lacks the contextual understanding of app structure.

The absence of UI exploration also impacts the LLM's ability to resolve ambiguities in S2R instructions. Without app-specific knowledge to contextualize the replay process, BugRepro struggles with tasks requiring nuanced understanding of app behavior, such as determining the correct sequence of actions across multiple screens. The slightly reduced LLM response time of 5.64s (compared to 9.37s) indicates less complex processing but at the expense of reproduction success.

The results confirm that RAG and UI exploration are not only individually essential but also synergistic in enabling BugRepro to achieve optimal performance. RAG enhances the LLM's understanding during S2R extraction by providing contextually relevant examples, while UI exploration complements this by guiding the replay process with app-specific knowledge. Together, these components address the diverse challenges inherent in bug reproduction, including ambiguous instructions, dynamic app states, and complex workflows, even though they increase the overall processing time to 124.7 seconds and LLM response time to 9.37 seconds.

\fbox{%
    \parbox{0.95\linewidth}{%
        \textbf{RQ3 summary:} The ablation study demonstrates the critical contributions of RAG and UI exploration to BugRepro's performance. RAG significantly enhances S2R extraction accuracy by providing task-relevant examples, while UI exploration enables precise and efficient navigation of complex app states. The synergistic relationship between these components allows BugRepro to address diverse challenges in bug reproduction tasks, establishing their necessity for achieving state-of-the-art results.
    }%
}

\subsection{RQ4: How does the robustness of each method vary across different language models?}\label{sec5.4}
To answer RQ4, we evaluate BugRepro's performance using both GPT-4 and our default DeepSeek model to assess its generalization capability. Table~\ref{tab:results} presents the results, demonstrating consistent S2R entity extraction performance across both models, with DeepSeek only showing slight disadvantages in Step (87.85\% vs. 87.61\%) and Component (33.87\% vs. 33.76\%) extraction. For replay performance, DeepSeek achieves a higher NSR of 96 compared to GPT-4's 86, though with a longer average reproduction time (124.7s vs. 83.0s).

\begin{table}[t]
\centering
\caption{Results Across Different LLMs}
\label{tab:results}
\begin{tabular*}{\columnwidth}{@{\extracolsep\fill}lrrrrr}
\toprule
\multicolumn{6}{c}{S2R Entity Extraction} \\
\midrule
Method & Step & Action & Component & Input & Direction \\
\midrule
Ours (gpt-4)    & 87.61\% & \textbf{69.60\%} & 33.76\% & \textbf{72.16\%} & \textbf{83.04\%} \\
Ours (deepseek) & \textbf{87.85\%} & 69.49\% & \textbf{33.87\%} & 70.93\% & 82.91\% \\
\midrule
\multicolumn{6}{c}{Replay Performance} \\
\midrule
Method & \multicolumn{1}{c}{NSR} & \multicolumn{2}{c}{Average Time} & \multicolumn{2}{c}{Average Response Time} \\
\midrule
Ours (gpt-4)    & 86 & \multicolumn{2}{c}{\textbf{83.0s}} & \multicolumn{2}{c}{\textbf{8.5s}} \\
Ours (deepseek) & \textbf{96} & \multicolumn{2}{c}{124.7s} & \multicolumn{2}{c}{9.3s} \\
\bottomrule
\end{tabular*}
\end{table}

These findings highlight BugRepro's model-agnostic architecture, which integrates RAG and UI exploration to enhance LLM capabilities. By transforming generic LLMs into specialized bug reproduction tools, BugRepro reduces dependencies on specific model features, enabling robust performance across different LLMs.

Through detailed analysis, we observe that while both models show comparable S2R extraction accuracy, their performance differs in key areas. DeepSeek demonstrates superior bug reproduction capability with higher NSR, though it requires more processing time, which is reflected in its slightly higher average response time (9.3s vs. 8.5s).
This also indicates that GPT-4 is efficient in rapid debugging scenarios, but its lower NSR indicates that it may encounter difficulties when handling certain complex application interactions.
In contrast, DeepSeek's higher success rate in bug reproduction indicates better performance in translating extracted S2R entities into effective app interactions, especially for complex scenarios involving multiple steps or conditional logic.

\fbox{%
\parbox{0.95\linewidth}{%
\textbf{RQ4 summary:} Results demonstrate BugRepro's ability to generalize across LLMs, with DeepSeek achieving higher reproduction success (96 vs. 86) at the cost of longer execution time (124.7s vs. 83.0s). This flexibility highlights BugRepro's adaptability to diverse LLMs and suggests opportunities for hybrid approaches that leverage the strengths of different models.
}%
}

\section{Threats to Validity}\label{sec6}
\textbf{The structured-output reliability threat} mainly lies in the selection of evaluation metrics and the inherent randomness in LLM outputs. LLMs frequently produce unstructured responses, which impedes their integration with structured code and automation pipelines. To address these challenges, we initially adopt approaches from prior work~\citep{adbgpt}, which attempt to enforce output formatting through prompt examples (e.g., ``1. [Tap] [‘bookmark’]''). However, such efforts have not fully standardized the responses. As illustrated in Table~\ref{tab:unstructured_llm_outputs_prior_work}, LLM outputs often display inconsistencies in format, structure, and verbosity, thereby complicating automation and diminishing accuracy.

\textbf{To confront this challenge,} our approach employs LangChain to enforce output constraints and ensure standardized formatting. The standardized outputs are shown in Table~\ref{tab:json-table}. The refinement of the output format further boosts the performance of BugRepro, enforces structured action output an ensure consistency and automation readiness.

\begin{table}[h]
\caption{Examples of unstructured LLM outputs from prior work}
\label{tab:unstructured_llm_outputs_prior_work}
\centering
\begin{tabular*}{\textwidth}{@{\extracolsep{\fill}}|l|l|}
\hline
Example output 1: & 
  \begin{tabular}[t]{@{}l@{}}
  \#\#\# Suggestion: \\
  1. \textbf{Tap on the ``share dialog'' widget}. \\
  2. \textbf{Switch apps}.
  \end{tabular} \\ \hline
Example output 2: & 
  \begin{tabular}[t]{@{}l@{}}
  \textbf{Suggestion:} \\
  1. Install the app ``org.mozilla.rocket.debug.ting'' if not already installed. \\
  2. Launch the app after installation.
  \end{tabular} \\ \hline
Example output 3: & 
  \begin{tabular}[t]{@{}l@{}}
  \textbf{Suggestion:} \\
  1. \textbf{Action:} Click the ``Skip''\textgreater{}\textgreater{} button to proceed. \\
    - \textbf{Feature:} ``Skip'' \textgreater{}\textgreater{}' \\
    - \textbf{Action:} ``click''
  \end{tabular} \\ \hline
Example output 4: & 
  \begin{tabular}[t]{@{}l@{}}
  \textbf{Suggestion:} \\
  1. \textbf{Explore the ``Licenses'' section}: Check for relevant options. \\
    {[}\{``action'': ``click'', ``feature'': ``Licenses''\}{]} \\
  2. \textbf{Scroll down} if no options are found. \\
    {[}\{``action'': ``scroll'', ``target\_direction'': ``down''\}{]} \\
  3. \textbf{Back to main screen} if needed. \\
    {[}\{``action'': ``back''\}{]}
  \end{tabular} \\ \hline
Example output 5: & 
  \begin{tabular}[t]{@{}l@{}}
  \#\#\# Suggestion: \\
  1. \textbf{Click on the ``LOG IN'' button}. \\
    - Action: {click} \\
    - Feature: {LOG IN} \\
  \end{tabular} \\ \hline
\end{tabular*}
\end{table}

\textbf{The internal threat} mainly lies in potential bugs in our implementations. \textbf{To mitigate this,} we leverage the provided artifacts for ReCDroid and AdbGPT to ensure correctness. For BugRepro, two authors independently review and rigorously test the code. 

\textbf{The external threat} mainly lies in the generalizability of the dataset. \textbf{In our experiments,} we collect all released bug reports from existing works and all of them are real-world bugs, which can be representative to some degree. However, the effectiveness of BugRepro on a wider range remains to be evaluated in the future.
Moreover, the conclusions derived from using a specific LLM (e.g., DeepSeek) may not generalize to other LLMs. To address this, we conduct additional experiments with GPT-4, an advanced LLM, and observe promising and consistent results.

\section{Related Work}\label{sec7}
\subsection{Traditional Android Bug Reproduction.}

Existing traditional Android bug reproduction technologies such as Yakusu~\citep{fazzini2018automatically} and ReCDroid~\citep{zhao2019recdroid}, Yakusu, the first method to propose integrating bug reports with bug reproduction, consists of three modules: ontology extraction, bug report analysis, and executable action search. The ontology extraction module analyzes application UI-related files to build an ontology that supports mapping. The bug report analysis module uses natural language processing techniques~\citep{mikolov2013distributed} to extract abstract steps and address logical gaps. The UI action search module takes abstract steps as input, dynamically searches and generates test cases, employing a depth-first search strategy~\citep{choi2013guided} to handle different abstract steps. Each module leverages different technical tools to achieve its functionality and generate specific format outputs.
ReCDroid, consists of two main phases: crash report analysis and dynamic exploration. The former uses 22 predefined grammar patterns and deep-learning NLP techniques~\citep{honnibal2017spacy} to map sentences into semantic representations. The latter phase is based on DOET and DFS, combined with Q-learning~\citep{zhao2019recdroid} to optimize matching and complete steps, generating the shortest event sequence. Existing approaches for integrating bug reports with reproduction processes suffer from several limitations. Many rely heavily on UI elements and general-purpose word embedding models, which restrict their ability to capture application-specific semantics and contextual nuances. Additionally, methods based on predefined grammar patterns often struggle to accurately interpret diverse and dynamic GUI components, leading to poor generalization and limited adaptability across different applications.

\subsection{Multimodal Bug Reproduction.}
To bridge the gap between language parsing and practical application components, a new method called ScopeDroid, based on multimodal processing~\citep{huang2023context,fang2021read,yang2017anserini}, is proposed.
ScopeDroid consists of three modules: STG (State Transition Graph) construction and information extraction, fuzzy reproduction step matching~\citep{liu2016ssd,robertson1995okapi}, and path planning with STG expansion. During construction, Droidbot~\citep{droidbot} is used to build the STG and extract component information. In the matching phase, a multimodal network structure is employed to calculate matching scores. In the path planning and expansion phase, the algorithm selects the optimal path~\citep{huang2023context} for execution or guided exploration. Additionally, since the STG may be incomplete, newly encountered components are incorporated, and the matrix is updated to re-plan the exploration.
ScopeDroid has the following limitations: it only considers the context of the current UI interface, cannot perform joint context matching for the entire report, and assumes that a single interaction between reproduction steps and GUI components may not hold, making it incapable of understanding steps that require reasoning or specific prior knowledge.

Compared to these methods, our approach has unique advantages. It leverages the exceptional natural language understanding capabilities of LLMs while closely integrating the context of bug reports and detailed information from the application interface, thereby providing a more effective and comprehensive solution to the related issues.

\subsection{LLM-based Android Bug Reproduction.}
LLMs demonstrate excellent performance in natural language processing, which leads to the proposal of the method AdbGPT. AdbGPT entails the manual creation of fifteen BRs~\citep{adbgpt} annotated with labels, which are utilized as exemplars. During subsequent processing, for each BR, the most appropriate 1 to 3 exemplars are selected to enable few-shot learning~\citep{wang2020generalizing} in large-scale LLMs, which helps improve accuracy.

ReBL adapts general-purpose LLMs for bug reproduction. It uses activity names for context and groups UI widgets to enhance understanding. It handles single/multiple actions, provides feedback (e.g., execution status, sequence detection), and employs summarization to manage token limits. Success is judged by linking error symptoms to UI info; exploration stops if reproduction fails.

However, during the experimental process, we find that these two methods are highly dependent on specific LLM models, and changing the model has a significant impact on performance.
At the same time, while LLMs are currently at the forefront of text processing BugRepro, their limitations are widely documented by numerous scholars ~\citep{wang2020generalizing,wei2022chain,martino2023knowledge,ji2023survey}. Among these, the most critical challenge arises when applying LLMs to the S2R and reproduction approach stages, particularly due to the ``hallucination'' ~\citep{martino2023knowledge,bang2023multitask,chang2024survey} problem. As the existing LLMs are not specifically trained to adapt to the analysis of bug reports, meanwhile, it is well-known that when posing questions to LLMs, the likelihood of generating ''hallucination'' responses increases in the absence of sufficient prior knowledge in the corresponding domain ~\citep{sahoo2024systematic,gao2023retrieval,just2018comparing,varshney2023stitch,wu2024faithful,zhang2022automatic}, leading to inaccurate outputs during the S2R stage, which causes the subsequent reproduction approach to function incorrectly, and the hallucination issue may further lead to erroneous exploration in the reproduction process. Addressing these challenges is essential to ensure reliable and accurate outputs.
The RAG approach demonstrates superior performance in enhancing the accuracy and reliability of generated content through its effective utilization of external knowledge. In contrast to methods~\citep{zhao2019recdroid} that rely exclusively on pre-trained language models, RAG operates by: (1) retrieving relevant text fragments from a knowledge repository in response to user queries, and (2) incorporating these fragments into a language generation model (e.g., GPT) to produce the final output. This combined retrieval-generation mechanism provides the language model with more precise and contextually relevant information, thereby: (i) substantially mitigating the risk of ``hallucinations'' (instances where models generate factually incorrect content), and (ii) consistently improving the overall quality of generated outputs.
By integrating existing RAG technology into the method and striving to provide the LLM with more context information closely related to the app under test, we successfully enhance the robustness of the method when switching between different LLMs.

In the process of automatically reproducing bugs based on bug report descriptions, previous methods have exhibited significant limitations. Our approach addresses these shortcomings through targeted improvements to existing techniques, resulting in substantial performance gains. 
\section{Conclusion}\label{sec8}
In conclusion, this paper proposes a novel method, named BugRepro, to address the challenges of unclear or incomplete S2Rs in bug reports and the inherent complexity of modern apps. By integrating domain-specific knowledge and leveraging a RAG method alongside app-specific GUI exploration, BugRepro enhances the accuracy, success rate, and efficiency of bug reproduction.
Extensive experimental results validate BugRepro’s superiority over state-of-the-art methods, achieving significant improvements across multiple metrics. This highlights BugRepro’s potential to streamline debugging processes and maintain high-quality user experiences in mobile application development.

In the future, BugRepro could be extended along several promising research directions to further advance its capabilities. First, the system’s contextual understanding could be improved by incorporating developer feedback patterns, enabling more precise interpretation of ambiguous S2Rs and enhancing reproduction accuracy. Second, a self-learning mechanism could be developed to iteratively refine reproduction strategies based on successful outcomes, thereby improving long-term performance. Third, extending the system’s cross-platform compatibility would allow bug reproduction across diverse operating systems and device configurations, broadening its practical utility. Fourth, integrating predictive bug analysis could enable proactive issue detection, potentially identifying defects before they are formally reported. Finally, an enhanced visualization system for reproduction steps could facilitate clearer communication between QA teams and developers, streamlining the debugging process. These extensions would collectively improve BugRepro’s robustness, generalizability, and usability in real-world development environments.

\textbf{Data availability}: The whole datasets generated and analyzed during the current study are available from the corresponding author on reasonable request. Here is a website that provides key code implementations and original datasets: \citep{anonymous_github_2025}.





\bibliography{sn-bibliography}

\end{document}